\documentclass[aps,prd,twocolumn,superscriptaddress,longbibliography,nofootinbib]{revtex4-2}
\usepackage[utf8]{inputenc}
\usepackage{color}
\usepackage{graphicx}

\usepackage{amsmath,amssymb,amsfonts}

\def\prl{Phys. Rev. Lett.}
\def\prd{Phys. Rev. D}

\def\apj{Astrophys. J.}

\def\apjl{Astrophys. J. Lett.}
\def\aap{Astronomy and Astrophysics}
\def\pr{Phys. Rev.}

\def\mnras{Mon. Not. R. Astron. Soc.}

\def\apss{Astrophys. Space Sci.}
\def\nphysa{Nucl. Phys. A}
\def\jcap{Journal of Cosmology and Astroparticle Physics}

\newcommand\longcomment[1]{}

\def\cm{\mbox{cm}}
\def\g{\mbox{g}}

\begin{document}

\title{Accretion onto black holes inside neutron stars with piecewise-polytropic equations of state: analytic and numerical treatments}

\author{Sophia C.~Schnauck}

\affiliation{Department of Physics and Astronomy, Bowdoin College, Brunswick, ME 04011, USA}

\author{Thomas W.~Baumgarte}

\affiliation{Department of Physics and Astronomy, Bowdoin College, Brunswick, ME 04011, USA}

\author{Stuart L.~Shapiro}

\affiliation{Departments of Physics and Astronomy, University of Illinois at Urbana-Champaign, Urbana, IL 61801, USA}

\begin{abstract}
We consider spherically symmetric accretion onto a small, possibly primordial, black hole residing at the center of a neutron star governed by a cold nuclear equation of state (EOS).  We generalize the relativistic Bondi solution for such EOSs, approximated by piecewise polytropes, and thereby obtain analytical expressions for the steady-state matter profiles and accretion rates.  We compare these rates with those found by time-dependent, general relativistic hydrodynamical simulations upon relaxation and find excellent agreement.  We consider several different candidate EOSs, neutron star masses and central densities and find that the accretion rates vary only little, resulting in an accretion rate that depends primarily on the black hole mass, and only weakly on the properties of the neutron star.
\end{abstract}

\maketitle

\section{Introduction}
\label{sec:intro}

Multiple authors have suggested that neutron stars may act as dark-matter detectors (e.g.~\cite{GolN89,deLF10,BraL14,BraE15,CapPT13,BraLT18,EasL19,GenST20}). In one scenario, primordial black holes (PBHs), which may either contribute to or even account for the dark matter in the Universe, may be captured by neutron stars and subsequently accrete the entire star \cite{Haw71,Mar95}.  The observed existence of neutron star populations has been invoked to constrain primordial black holes in a mass window of about $10^{-15} M_{\odot} \lesssim M_{\rm BH} \lesssim 10^{-9} M_{\odot}$ (see \cite{CapPT13}) that is poorly constrained by other arguments and observations (see, e.g.~\cite{KueF17,CarK20,CarKSY20,SasSTY18,VasV21}).  Even this constraint assumes that both the capture and accretion processes are sufficiently fast.  While different authors have arrived at different estimates for the rates of the former \cite{CapPT13,GenST20,MonFVSH19,KaiNSY21}, we will derive analytical rates for the latter in this paper, even for neutron stars governed by realistic nuclear equations of state (EOSs).

In an alternative scenario, other candidate dark-matter particles, possibly including axions, also can be captured by neutron stars.  Under sufficiently favorable conditions, these particles may coalesce to form a high-density object that then collapses to a small black hole (e.g.~\cite{GolN89,deLF10,BraL14,BraLT18}), thereby resulting in the same accretion process as the scenario above.  Independently of the precise scenario, it is of interest to explore the rate at which a central ``endoparasitic" black hole disrupts its host neutron star.

Spherically symmetric, steady-state accretion onto a point mass of a fluid that is homogeneous and at rest far from the mass is described by the Bondi solution in Newtonian physics \cite{Bon52}.  Most treatments of Bondi accretion, or its relativistic counterpart describing accretion onto a Schwarzschild black hole \cite{Mic72} (see also Appendix G in \cite{ShaT83}, hereafter ST), focus on soft EOSs with adiabatic indices $1 \leq \Gamma \leq 5/3$, which is suitable for most astrophysical plasmas.  While the EOS governing the cores of neutron stars is not known, most realistic candidates for the EOS at nuclear densities are stiff.  As we discussed in \cite{RicBS21a} (hereafter RBS), accretion for stiff EOSs with $\Gamma > 5/3$ shows some qualitative differences from that of soft EOSs.  In particular, there exists a {\em minimum} steady-state accretion rate for stiff EOSs.  As shown in Appendix A of \cite{RicBS21b}, these results also hold for a black hole inside a neutron star, as long as the black hole mass $M_{\rm BH}$ is much smaller than that of the neutron star, $M_{\rm BH} \ll M_{\rm NS}$. In this case the relativistic Bondi formalism yields the accretion rate as measured by a ``local asymptotic observer", i.e.~one who is far from the black hole, but deep inside the neutron star.  The above minimum accretion rate therefore results in a {\em maximum survival time} for neutron stars harboring a black hole \cite{BauS21}.

A number of authors have also performed numerical simulations of accretion onto endoparasitic black holes inside neutron stars.  East and Lehner \cite{EasL19} considered black holes with masses $M_{\rm BH} \geq 10^{-2} M_{\rm NS}$ and adopted nuclear EOSs (modeled with the ``piecewise-polytrope" approximation described below).  In particular, they observed that the accretion rate is proportional to $M_{\rm BH}^2$, as expected from the relativistic Bondi formalism (see eq.~\ref{m_dot} below), and that the effects of rotation are small (see also \cite{KouT14}).  Focusing on nonrotating configurations and stiff $\Gamma$-law EOSs, we performed simulations for much smaller black holes with masses $M_{\rm BH} \gtrsim 10^{-9} M_{\rm NS}$, and found that the accretion rates agree very well with those predicted by the Bondi formalism, even quantitatively \cite{RicBS21b}.

In this paper we generalize the results of RBS and \cite{RicBS21b} to allow for realistic, nuclear EOSs, approximated by piecewise polytropes (PWPs).   We review the PWP treatment of EOSs in Section \ref{sec:eos}, and then derive analytical expressions for the stationary accretion rates onto small black holes inside neutron stars governed by such EOSs in Section \ref{sec:bondi}.  We perform  time-dependent, numerical simulations in full general relativity for these EOSs as described in Section \ref{sec:numerics}.  Once these numerical solutions have relaxed into a quasi-stationary solution they agree very well with the analytical solutions, as shown in Section \ref{sec:results}.   Briefly summarizing in Section \ref{sec:summary}, we find that, for these realistic nuclear EOSs,  the accretion rates depend only weakly on the EOS and the neutron star's density, and hence mainly on the black hole mass. These rates are just slightly larger than the minimum accretion rates that RBS and \cite{BauS21} computed under the assumption of $\Gamma$-law EOSs.  Unless noted otherwise we use geometrized units with $G = 1 = c$ in this paper, and adopt the solar mass $1 M_\odot = 1.99 \times 10^{33} \, \mbox{g} = 1.47 \times 10^5 \, \mbox{cm} = 4.9 \times 10^{-6} \, \mbox{s}$ as the fundamental unit.

\section{Nuclear EOSs: approximation by piecewise polytropes}
\label{sec:eos}

\begin{table}[t]
    \centering
    \begin{tabular}{c||c|c}
         & $\rho_0 \, [\mbox{g cm}^{-3}]$ & $\rho_0 \, [M_\odot^{-2}]$  \\
         \hline
         $\rho_{0,1}$ & ~$1.46 \times 10^{14}$~ & ~$2.37 \times 10^{-4}$~\\
         $\rho_{0,2}$ & $5.01 \times 10^{14}$ & $8.11 \times 10^{-4}$ \\
         $\rho_{0,3}$ & $1.00 \times 10^{15}$ & $1.62 \times 10^{-3}$ 
    \end{tabular}
    \caption{Values of the ``boundary" rest-mass densities that separate the different regions in the PWP approach.  For each $i$, the constants $\Gamma_i$, $K_i$, and $b_i$ listed in Table \ref{table:PWP} apply between the densities $\rho_{0,i}$ and $\rho_{0,i+1}$.  Values for the rest-mass density $\rho_0$ in cgs units, $\rho_0^{\rm cgs}$, are related to those in units of solar masses, $\rho_0^{M_\odot}$, by $\rho_0^{\rm cgs} = \rho_0^{M_\odot} c^2 \,G^{-1} (M_\odot / 1.47 \times 10^5 \, \mbox{cm})^{2}$.
    }
    \label{tab:rho_boundaries}
\end{table}

\begin{table*}[t]
\centering
\begin{tabular}{c||c|c|c||c|c|c|c||c|c|c||c|c}
     EOS & $\Gamma_1$ & $\Gamma_2$ & $\Gamma_3$ & $K_0$ & $K_1$ & $K_2$ & $K_3$ & $b_1$ & $b_2$ & $b_3$ & $M_{\rm max} \, [M_{\odot}]$ & $\rho_{0c}^{\rm max} [\g \, \cm^{-3}]$  \\
    \hline
    SLy & 3.005 & 2.988 & 2.851 & 0.089492 & 84572.6 & 74935.5 & 31071.4 & 0.0104711 & 0.0102419 & 0.00234129 & 2.06 & $2.01\times 10^{15}$
    \\
    AP3 & 3.166 & 3.573 & 3.281 & 0.0747568 & 270916 & 4906280 & 751390 & 0.00888779 & 0.0128859 & –0.00322345 & 2.38 & $1.67 \times 10^{15}$
    \\
    AP4 & 2.83 & 3.445 & 3.348 & 0.0851938 & 18679.1 & 1486420 & 796972 & 0.00976174 & 0.0154307 & 0.011658 & 2.20 & $1.92 \times 10^{15}$
    \\
    MS1 & 3.224 & 3.033 & 1.325 & 0.2032526 & 1970000 & 307447 & 5.26309 & 0.0243216 & 0.0175589 & –1.66799 & 2.74 & $1.08 \times 10^{15}$
    \\
    H4  & 2.909 & 2.246 & 2.144 & 0.194153 & 82323.1 & 735.161 & 381.707 & 0.0224696 & -0.00640722 & -0.0239386 & 2.00 & $1.70 \times 10^{15}$  \\
\end{tabular}
\caption{List of EOSs considered in this paper together with their PWP parameters.  The adiabatic coefficients $\Gamma_i$ as well as the constants $b_i$ are dimensionless, while, in geometrized units, the coefficients $K_i$ have units of length (or mass) to the power $2 (\Gamma_i - 1)$, which we express in units of solar mass.  We use $\Gamma_0 = 1.35692$ in the lowest density region for all EOSs, adopting the value for the highest-density crust piece in \cite{ReaLOF09} (see their Table II).  We also provide the maximum gravitational mass $M_{\rm max}$ of non-rotating neutron stars, and the corresponding central density $\rho_{0c}^{\rm max}$.}
\label{table:PWP}
\end{table*}

As demonstrated by \cite{ReaLOF09}, a large class of candidates for realistic, cold nuclear EOSs can be approximated remarkably well with {\em piecewise polytropes} (PWPs).  Specifically, we write the pressure $P$ as a function of the rest-mass density $\rho_0$ as 
\begin{equation} \label{PWP:P}
    P = K_i \rho_0^{\Gamma_i}
\end{equation}
where the constants $K_i$ and $\Gamma_i$ take different values in four different density ``regions" labeled by the index $0 \leq i \leq 3$.  The different regions are separated by three ``boundary densities" that can be chosen to be the same for all EOSs (see Table \ref{tab:rho_boundaries}).  We follow \cite{TsoRS20} in our implementation of the PWP EOS; in particular, we model the low-density crust with only one piece rather than the four pieces adopted by \cite{ReaLOF09} (see their Table II), and we also choose the values for the boundary densities as in \cite{TsoRS20}.  The different values $\Gamma_i$, as well as one of the constants $K_i$, can be found from fits to each EOS; the remaining values of $K_i$ are then related to each other by imposing continuity of $P$ across the boundaries between different regions. The specific internal energy $\epsilon$ then takes the form
\begin{equation} \label{PWP:eps}
    \epsilon = b_i + \frac{K_i}{\Gamma_i - 1} \rho_0^{\Gamma_i - 1}
\end{equation}
(see eq.~5 in \cite{ReaLOF09}) where the $b_i$ are constants of integration that are chosen to make $\epsilon$ continuous across the boundaries between different regions.  In the lowest-density region the constant $b_0$ vanishes, but in general $b_i \ne 0$ for $i > 0$.\footnote{Note that \cite{ReaLOF09} used symbols $a_i$ for these constants; we choose $b_i$ here in order to avoid confusion with the sound speed $a$ below.}  The total mass-energy density $\rho$ is then
\begin{equation}
    \rho = \rho_0 ( 1 + \epsilon) 
    = \rho_0 + b_i \rho_0 + \frac{K_i}{\Gamma_i - 1} \rho_0^{\Gamma_i}.
\end{equation}
While $P$, $\epsilon$ and $\rho$ are continuous across density boundaries, the speed of sound $a$, computed from
\begin{align} \label{a_of_rho}
    a^2 & = \left( \frac{dP}{d \rho} \right)_s
    = \frac{\rho_0}{\rho + P} \left( \frac{dP}{d\rho_0} \right)_s \nonumber \\
    & = \frac{\Gamma_i K_i \rho_0^{\Gamma_i - 1}}{1 + b_i + \Gamma_i K_i \rho_0^{\Gamma_i - 1}/(\Gamma_i - 1)},
\end{align}
where the subscript $s$ denotes constant entropy, in general is not continuous -- in fact, $a$ may not even grow monotonically with $\rho_0$.  In each region $i$ we may invert (\ref{a_of_rho}) to find
\begin{equation} \label{rho_0_1}
    \Gamma_i K_i \rho_0^{\Gamma_i - 1} = 
    \frac{a^2\, (1 + b_i)}{1 - a^2 / (\Gamma_i - 1)},
\end{equation}
but, since $a$ may neither be a continuous nor a monotonic function of $\rho_0$, we may not be able to invert (\ref{a_of_rho}) globally.

A list of the $\Gamma_i$ for a large number of EOSs is given in Table III of \cite{ReaLOF09}.  In this paper we consider representatives  of four different families of EOSs, namely the SLy \cite{DouH01}, AP3 and AP4 \cite{AkmPR98}, MS1 \cite{MueS96}, and H4 \cite{LacNO06} EOSs, which represent different theoretical approaches to constructing realistic nuclear EOSs.  For these EOSs we provide all the above PWP parameters  in Table \ref{table:PWP}.  All these EOSs result in maximum allowed masses that are consistent with all observed neutron star masses, including the largest currently known mass of $2.08^{+0.07}_{-0.07} M_\odot$ (where the errors represent a 68.3\% credibility interval), reported for the millisecond pulsar J0740+6620 from measurements of the relativistic Shapiro effects (see \cite{Croetal20,Fonetal21}; see also \cite{Riletal21} for NICER and XMM analysis of the same pulsar, resulting in a consistent value for its mass, as well as \cite{Miletal21} for constraints on the EOS resulting from NICER radius measurements of J0740+6620).   Other high-mass neutron stars include PSR J1614-2230 with a mass of $M = 1.93 M_{\odot}$ (see \cite{Fonetal16}), and PSR J0348+0432 with a mass of $M = 2.01 M_{\odot}$ (see \cite{Antetal13}).  With the exception of MS1 and H4, the above EOSs are also consistent with the neutron star masses and tidal distortions inferred from the gravitational wave signal GW170817 and electromagnetic follow-up observations \cite{Abbetal19}; while both MS1 and H4 appear to be ruled out based on tidal distortions, we include them regardless as examples of stiff EOSs.

\section{Bondi accretion for piecewise polytropes: analytical solution}
\label{sec:bondi}

We now generalize the Bondi solution \cite{Bon52,Mic72,ShaT83}, describing stationary, adiabatic, spherically symmetric fluid flow onto a Schwarzschild black hole, for PWPs.  We follow the derivation in Appendix G of ST up to their eq.~(G.22) (hereafter ST.G.22) unchanged, since it does not yet make any assumptions about the EOS.  In particular (ST.G.21), the integrated continuity equation
\begin{equation} \label{continuity}
    4 \pi \rho_0 u r^2 = \mbox{const},
\end{equation}
as well as (ST.G.22), the integrated Euler equation
\begin{equation} \label{bernoulli}
    \left( \frac{ \rho + P}{\rho_0} \right)^2 
    \left( 1 - \frac{2 M_{\rm BH} }{r} + u^2 \right) = \mbox{const},
\end{equation}
remain valid, as does (ST.G.17) for the relations at the critical radius,
\begin{equation} \label{crit_condition}
    u_s^2 = \frac{a_s^2}{1 + 3 a_s^2} = \frac{M_{\rm BH}}{2 r_s},
\end{equation}
(provided this critical radius exists; see our discussion below).  In the above equations $r$ is the areal radius and $u \equiv |u^r|$ the {\em inward} radial component of the fluid's four-velocity; note that the above expressions employ Schwarzschild coordinates.

Rather than adopting a single polytrope, as in ST and RBS, we now adopt the PWPs described above.  In particular, the first factor on the left-hand side of (\ref{bernoulli}) then takes the form
\begin{equation} \label{one_over_h_1}
    \frac{\rho + P}{\rho_0} = 1 + b_i + \frac{\Gamma_i}{\Gamma_i - 1} K_i \rho_0^{\Gamma_i - 1}.
\end{equation}
Note also that, using (\ref{rho_0_1}), we may relate the density at the critical point to that in the ``local asymptotic region", denoted by a subscript $*$, by
\begin{equation}
    \rho_{0s}^{\Gamma_s - 1} = \frac{\Gamma_* K_*}{\Gamma_s K_s} \, \frac{\Gamma_* - 1 - a_*^2}{\Gamma_s - 1 - a_s^2} \,
    \frac{\Gamma_s - 1}{\Gamma_* - 1} \, \frac{1 + b_s}{1 + b_*} \, \frac{a_s^2}{a_*^2} \rho_{0*}^{\Gamma_* - 1}
\end{equation}
({\it cf.}~eq.~10 in RBS, hereafter RBS.10).  Inserting this into (\ref{continuity}), and using (\ref{crit_condition}), we may now write the accretion rate as observed by a local asymptotic observer (hence the superscript $*$) as
\begin{equation} \label{m_dot}
    \dot M_0^* = 4 \pi \lambda \left( \frac{M_{\rm BH}}{a_*^2} \right)^2 \rho_{0*} a_*,
\end{equation}
which assumes the same form as, e.g., (ST.G.33) or (RBS.11), except that the dimensionless accretion eigenvalue $\lambda$ is now given by
\begin{align} \label{lambda}
    \lambda = & \left( \frac{a_s}{a_*} \right)^{(5 - 3 \Gamma_s)/(\Gamma_s - 1)} \rho_{0*}^{(\Gamma_* - \Gamma_s)/(\Gamma_s - 1)} \frac{(1 + 3 a_*^2)^{3/2}}{4}  \nonumber \\
    & \left( \frac{\Gamma_* K_*}{\Gamma_s K_s} \, \frac{\Gamma_* - 1 - a_*^2}{\Gamma_s - 1 - a_s^2} \, 
    \frac{\Gamma_s - 1}{\Gamma_* - 1} \, \frac{1 + b_s}{1 + b_*} \right)^{1/(\Gamma_s - 1)}.
\end{align}
Note that this reduces to eq.~(RBS.12), as expected, when the critical point is in the same density region as the asymptotic observer, so that $\Gamma_s = \Gamma_*$, $K_s = K_*$ and $b_s = b_*$.  Note also that $M_{\rm BH}$ on the right-hand side of (\ref{m_dot}) is the black hole's gravitational mass, while the left-hand side is the rate at which rest mass crosses the black hole's horizon.

In order to evaluate $\lambda$ for given asymptotic values we need to relate $a_s$ to $a_*$,  which we will do using eq.~(\ref{bernoulli}).  We start by inserting (\ref{rho_0_1}) into (\ref{one_over_h_1}) to obtain
\begin{equation}
    \frac{\rho + P}{\rho_0} = (1 + b_i) \left( 1 + \frac{a^2}{\Gamma_i - 1 - a^2} \right).
\end{equation}
Evaluating the left-hand side of (\ref{bernoulli}) both at $r_s$ and in the local asymptotic region, where $r_* \gg M$ and $u_* \ll 1$, then yields
\begin{align}
    & \left(1 - \frac{2M}{r_s} + u_s^2 \right) (1 + b_s)^2 \left(1 + \frac{a_s^2}{\Gamma_s - 1 - a_s^2} \right)^2 \nonumber \\ 
    & ~~~~ = (1 + b_*)^2 \left(1 + \frac{a_*^2}{\Gamma_* - 1 - a_s^2} \right)^2
\end{align}
({\it cf.}~ST.G.29), or, taking the inverse of both sides and using (\ref{crit_condition}),
\begin{equation} \label{a_s}
    (1 + 3 a_s^2) \left(1 - \frac{a_s^2}{\Gamma_s - 1} \right)^2
    = \left( \frac{1 + b_s}{1 + b_*} \right)^2 \left(1 - \frac{a_*^2}{\Gamma_* - 1} \right)^2
\end{equation}
({\it cf.}~ST.G.30).  As for a single polytropic EOS, the relation (\ref{a_s}) forms a cubic equation for $x = a_s^2$ that we may write as
\begin{equation}
x^3 + A x^2 + B x + C = 0    
\end{equation}
with
\begin{align}
    A & = \frac{1}{3}(7 - 6 \Gamma_s) \nonumber \\
    B & = \frac{1}{3}(1 - \Gamma_s)(5 - 3 \Gamma_s) \\
    C & = \frac{(\Gamma_s - 1)^2}{3} \left( 1 - \left( \frac{1 + b_s}{1+b_*} \right)^2 \left(1 - \frac{a_*^2}{\Gamma_* - 1} \right)^2 \right)
    \nonumber
\end{align}
({\it cf.}~RBS.19).  Unlike in RBS, however, we now need to evaluate the constants $b_i$ and $\Gamma_i$ in the regions corresponding to $\rho_{0*}$ and $\rho_{0s}$ (or, equivalently, $a_*$ and $a_s$).  For a given value of $\rho_{0*}$, we know how to choose $b_*$ and $\Gamma_*$, but, unless the local asymptotic values are in the highest-density region $i = 3$ already, we do not know a priori in which density region the critical point will be.  Stated differently, solving (\ref{a_s}) for $a_s$ requires values $b_s$ and $\Gamma_s$, but choosing those depends on what region $a_s$ ends up in.  We can solve this problem as follows.  

\begin{figure}
    \centering
    \includegraphics[width = 0.45 \textwidth]{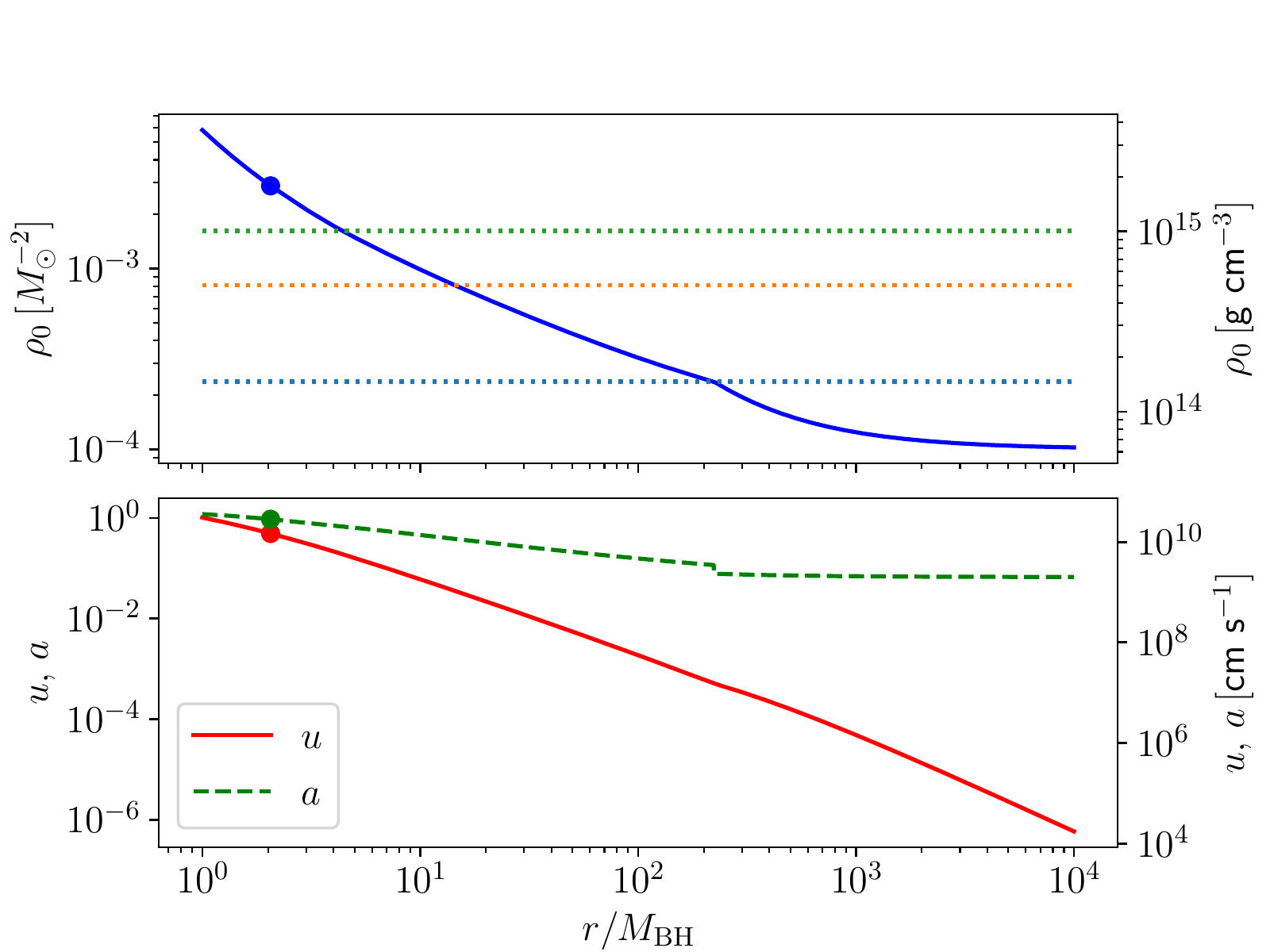}
    \caption{Analytical Bondi accretion profiles for the SLy EOS and an asymptotic density $\rho_{0*} = 10^{-4} M_{\odot}^{-2} = 6.2 \times 10^{13} \, \mbox{g cm}^{-3}$.  The critical (areal) radius $r_s = 2.057 \, M_{\rm BH}$ is marked by the dots.  The horizontal dotted lines in the top panel mark the boundaries between the four density regions (see Table \ref{tab:rho_boundaries}); note the small discontinuities in the sound speed $a$ at the corresponding locations.}
    \label{fig:sample_profile}
\end{figure}

Say $\rho_{0*}$ is in density region $j$.  Assuming that $\rho_{0s} \geq \rho_{0*}$ we then consider all regions $i \geq j$, and solve eq.~(\ref{a_s}), using Cardano's formula as described in RBS, to obtain candidate solutions $a_{si}$ for each one, disregarding unphysical solutions for which $a_{si}^2 < 0$. We then evaluate (\ref{rho_0_1}) for remaining solutions $a_{si}$ in region $i$, and keep only those candidate solutions for which the corresponding rest-mass density $\rho_{0i}$ is indeed in region $i$.  In some cases, for low values of $a_*$, we still find viable solutions in multiple regions from this procedure.  For each one of these remaining solutions we can then construct fluid profiles by integrating eqs.~(ST.G.10) both inwards and outwards away from the critical radius $r_{si}$.  For the examples that we considered, 
at most one solution $a_{si}$ resulted in global solutions with exactly one critical point.  In the following we always adopt this solution as the analytical Bondi profile.  We show an example of such a profile for the SLy EOS, extending over all four density regions, in Fig.~\ref{fig:sample_profile}.  The above approach reduces to the simpler single Gamma-law case treated in \cite{RicBS21b}, of course, if $\rho_{0*}$ is in the highest-density region already.

Also note that the above procedure is not guaranteed to yield solutions.  As demonstrated in Appendix G in ST, the existence of a critical radius {\em is} guaranteed if all functions are continuous.  Specifically, ST argue that, since their function
\begin{equation}
D = \frac{u^2 - (1 - 2 M_{\rm BH}/r + u^2) a^2}{u \rho_0}
\end{equation}
(see ST.G.13) is negative for large $r$ but positive for small $r$ approaching the black hole horizon, it must have a root, which then provides the condition (\ref{crit_condition}).  In our treatment here, $D$ must still change sign, but since, for piecewise polytropes, it can no longer be assumed to be continuous everywhere, this does not imply that it {\em necessarily} has a root.  For most examples that we considered we were able to find critical points without any problems.  For small densities for the MS1 EOS, however, the above procedure did not yield any solutions, which we believe is related to the large discontinuity in the sound speed resulting from the large difference between $\Gamma_2$ and $\Gamma_3$.   In Fig.~\ref{fig:crit_point} below we show numerical results demonstrating that, in this case, a discontinuity in the sound speed leads to a ``jump" across the critical point at which the last two terms in eq.~(\ref{crit_condition}) are equal, so that a strict root of $D$ does not exist.  Clearly, this behavior is an artifact of the piecewise-polytropic treatment of the EOS, which results in these discontinuities.

\section{Numerical treatment}
\label{sec:numerics}

We complement our analytical results by performing numerical simulations of the accretion onto black holes at the center of neutron stars as follows.  

We construct initial data from a solution to the Tolman-Oppenheimer-Volkoff equations \cite{Tol39,OppV39} for a given EOS and a given central density.  Following \cite{RicBS21b} we then adopt a generalized puncture method to place a black hole with puncture mass ${\mathcal M}$ at the center of the neutron star, and solve the Hamiltonian constraint for the conformal factor $\psi$, 
assuming a moment of time symmetry, to obtain solutions to Einstein's constraint equations.  As demonstrated in Section III.C.1 of \cite{RicBS21b}, for ${\mathcal M} \ll M_{\rm NS}$ the black hole's gravitational mass is well approximated by $M_{\rm BH} \simeq \psi_{\rm NS} \, {\mathcal M}$, where $\psi_{\rm NS}$ is the conformal factor at the center of the unperturbed neutron star.

We then evolve these data using the Baumgarte-Shapiro-Shibata-Nakamura (BSSN) formalism (\cite{NakOK87,ShiN95,BauS99}; see also \cite{BauS10} for a textbook discussion), implemented in spherical polar coordinates \cite{BauMCM13,BauMM15} with the help of a reference-metric formalism \cite{BonGGN04,ShiUF04,Bro09,Gou12}.  We adopt moving-puncture coordinates, i.e.~"1+log" slicing for the lapse \cite{BonMSS95} and a ``Gamma-driver" condition for the shift \cite{Alcetal03,ThiBB11}, starting with a ``pre-collapsed" lapse $\alpha = \psi^{-2}$ and vanishing shift.  We evolve the equations of relativistic hydrodynamics using a Harten-Lax-van-Leer-Einfeld approximate Riemann solver \cite{HarLL83,Ein88} together with a simple monotonized central-difference limiter reconstruction scheme \cite{Van77}.  

Even though we start with initial data describing cold fluids, we allow for heating (e.g, by shocks) by adding to the cold pressure (\ref{PWP:P}) thermal contributions. 
Specifically, we compute thermal contributions to the internal energy density $U \equiv \rho_0 \epsilon$ from
\begin{equation} \label{u_th}
    U_{\rm th} = U - U_{\rm cold} = \rho_0 ( \epsilon - \epsilon_{\rm cold}),
\end{equation}
where $\rho_0$ and $\epsilon$ are computed from the dynamically evolved quantities, and $\epsilon_{\rm cold}$ is given by (\ref{PWP:eps}).  We then write $U_{\rm th} = U_{\rm nucl} + U_{\rm rad}$, where $U_{\rm nucl}$ accounts for finite-temperature corrections to an ideal, nonrelativistic, nucleon Fermi gas,
\begin{equation} \label{u_nucl}
    U_{\rm nucl} = \frac{(3 \pi^2)^{1/3} m_{\rm B}}{6 \hbar^2} n^{1/3} (k_{\rm B} T)^2,
\end{equation}
and $U_{\rm rad}$ for contributions from radiation,
\begin{equation} \label{u_rad}
    U_{\rm rad} = \eta a_{\rm rad} T^4
\end{equation}
(compare, e.g., \cite{BauST95}).   In the above equations $m_{\rm B}$ is the baryon rest mass (which we take to be equal to the neutron rest mass), $n = \rho_0 / m_{\rm B}$ the baryon number density, $k_{\rm B}$ the Boltzmann constant, $\hbar$ Planck's constant,  $T$ the temperature, $a_{\rm rad}$ the radiation constant, and the non-dimensional constant $\eta$ depends on which particles contribute to the radiation.  Allowing for photons ($\eta_{\rm ph} = 1$), three flavors of neutrinos ($\eta_{\nu} =  3 \times 7/8$), as well as electron-positron pairs ($\eta_{\rm pairs} = 7/4$) we have $\eta = \eta_{\rm ph} + \eta_\nu + \eta_{\rm pairs} = 43/8$.  We insert (\ref{u_nucl}) and (\ref{u_rad}) into (\ref{u_th}) and use a root-finding method to find the temperature $T$.  Knowing $T$, we can finally compute the pressure $P = P_{\rm cold} + P_{\rm th}$ using
\begin{equation}
P_{\rm th} = P_{\rm nucl} + P_{\rm rad} =  (\Gamma_{\rm th} - 1 ) \, U_{\rm nucl} + \frac{1}{3} U_{\rm rad}
\end{equation}
where $\Gamma_{\rm th} = 5/3$ for nonrelativistic nucleons.

In all our simulations we observe a transition from our (astrophysically artificial) initial data to a nearly time-independent equilibrium solution describing accretion onto the black hole.  Especially for initial data with smaller initial densities we see that this transition launches an outgoing shock wave; the accretion solution is then attained inside this shock wave.   While this shock wave does lead to some heating, we find that this heating is small, especially for large initial densities, and presumably transient, so that we still find good agreement between our numerical and analytical solutions, even though the latter has been constructed for a cold gas.

In order to measure the accretion rate we monitor the flux ${\mathcal F}$ of rest mass through spheres ${\mathcal S}$ of radius $r$,
\begin{equation}
{\mathcal F}(r) = - \int_{\mathcal S} \sqrt{-g} \,\rho_0 u^r d\theta d \varphi,
\end{equation}
where $g$ is the determinant of the spacetime metric (see, e.g., Appendix A in \cite{FarLS10}).  The accretion rate is then given by the flux ${\mathcal F}$ evaluated on the horizon
\begin{equation} \label{m_dot_numerical}
    \dot M_0 = {\mathcal F}(r_{\rm hor}).
\end{equation}
As in (\ref{m_dot}), this rate measures the accretion of rest mass rather than the change of the black hole's gravitational mass.  Note also that (\ref{m_dot_numerical}) measures the accretion rate as seen by an observer at infinity, i.e.~at large distances $r \gg R_{\rm NS}$ from the neutron star, while (\ref{m_dot}) measures that as seen by a local asymptotic observer with $M_{\rm BH} \ll r \ll R_{\rm NS}$.  As discussed in Section III.C.2 of \cite{RicBS21b}, we may compute the former from the latter using
\begin{equation} \label{translate_m_dot}
    \dot M_0 = \alpha_* \dot M_0^*,
\end{equation}
where $\alpha_*$ is the lapse function in the local asymptotic region.

\section{Results}
\label{sec:results}

\begin{table*}[t]
    \centering
    \begin{tabular}{c||c|c|c|c|c|c|c|c|c|c}
        EOS & $\rho_{0*} \, [\mbox{g cm}^{-3}]$
        & $M_{\rm BH} \, [M_\odot]$ & $M \, [M_\odot]$ & $\lambda$ & $\dot M_0^* / M_{\rm BH}^2 \, [M_\odot^{-2}]$ & $\dot M_0^*$~\footnote{Values for accretion rates $\dot M_{0}$ in units of solar mass per year, $\dot M_{0} \, [M_\odot / \mbox{yr}]$, can be computed from the dimensionless values $\dot M_{0}$ provided here using $\dot M_{0} \, [ M_\odot / \mbox{yr} ] = 6.43 \times 10^{12} M_\odot \, \mbox{yr}^{-1} \dot M_{0}$.}  & $\alpha_*$ & $\alpha_* \dot M_0^{*~\rm a}$ & $\dot M_0^{~\rm a}$ & $\dot M_0 / M_{\rm BH}^2 \, [M_\odot^{-2}]$ \\
        \hline \hline
        SLy & $1.99 \times 10^{15}$ & $1.52 \times 10^{-6}$ & 2.06 & 3.41 & 0.139 & $3.21 \times 10^{-13}$ &     0.432 & $1.39 \times 10^{-13}$ & $1.39 \times 10^{-13}$ & 0.060 \\
        & $9.92 \times 10^{14}$  & $1.25 \times 10^{-6}$ & 1.56 & 1.55 & 0.0969 & $1.51 \times 10^{-13}$ &     0.636 & $9.60 \times 10^{-14}$ & $9.70 \times 10^{-14}$ & 0.062 \\
        & $4.96 \times 10^{14}$ & $1.08 \times 10^{-6}$ & 0.579 & 0.442 & 0.0812 & $9.47\times 10^{-14}$ &     0.853 & $8.08\times 10^{-14}$ & $8.09\times 10^{-14}$ & 0.069 \\
        \hline 
        AP3 & $1.68 \times 10^{15}$ & $1.58 \times 10^{-6}$ & 2.38 & 4.67 & 0.099 & $2.47 \times 10^{-13}$ &     0.398 & $9.83 \times 10^{-14}$ & $9.91 \times 10^{-14}$ & 0.040 \\
        & $8.37 \times 10^{14}$ & $1.25 \times 10^{-6}$ & 1.61 & 1.68 & 0.0682 & $1.07\times 10^{-13}$ &     0.645 & $6.90\times 10^{-14}$ & $6.83 \times 10^{-14}$ & 0.044\\
        & $4.19 \times 10^{14}$ & $1.06 \times 10^{-6}$ & 0.402 & 0.254 & 0.0595 & $6.69\times 10^{-14}$ &     0.890 & $5.95\times 10^{-14}$ & $5.95 \times 10^{-14}$ & 0.053 \\
        \hline
        AP4 & $1.80 \times 10^{15}$ & $1.55 \times 10^{-6}$ & 2.20 & 4.36 & 0.107 & $2.58 \times 10^{-13}$ &     0.418 & $1.08 \times 10^{-13}$ & $1.07\times 10^{-13}$ & 0.045  \\
        & $8.99 \times 10^{14}$ & $1.22 \times 10^{-6}$ & 1.37 & 1.35 & 0.0770 & $1.15 \times 10^{-13}$ &     0.675 & $7.76\times 10^{-14}$ & $7.71\times 10^{-14}$ & 0.052 \\
        & $4.50 \times 10^{14}$ & $1.05 \times 10^{-6}$ & 0.36 & 0.201 & 0.0686 & $7.56 \times 10^{-14}$ &     0.897 & $6.78 \times 10^{-14}$ & $6.80\times 10^{-14}$ & 0.062 \\
        \hline
        MS1 & $9.30 \times 10^{14}$  & $1.45 \times 10^{-6}$ & 2.69 & 6.57 & 0.141 & $2.96\times 10^{-13}$      & 0.474 & $1.40\times 10^{-13}$ & $1.42\times 10^{-13}$ & 0.068 \\
        & $4.65 \times 10^{14}$ & $ 1.18\times 10^{-6}$ & 1.54 &  &  &  & 0.717 &  & $5.38\times 10^{-14}$     & 0.039 \\
        & $2.33 \times 10^{14}$ & $ 1.05\times 10^{-6}$ & 0.39 &  &  &  & 0.910 &  & $4.35\times         10^{-14}$ & 0.039\\
        \hline
        H4  & $1.55 \times 10^{15}$ & $ 1.40\times 10^{-6}$ & 2.00 & 1.98 & 0.176 & $3.45\times 10^{-13}$ & 0.508 & $1.75\times 10^{-13}$  & $1.766\times 10^{-13}$  & 0.090 \\
            & $7.75 \times 10^{14}$ & $ 1.24\times 10^{-6}$ & 1.68 & 1.28 & 0.121 & $1.86\times 10^{-13}$ & 0.655 & $1.22\times 10^{-13}$ & $1.213\times 10^{-13}$ & 0.079
            \\
            & $3.88 \times 10^{14}$ & $ 1.10\times 10^{-6}$ & 0.81 & 0.772 & 0.0916 & $1.11\times 10^{-13}$ & 0.830 & $9.21\times 10^{-14}$ & $9.26\times 10^{-14}$ & 0.077\\
    \end{tabular}
    \caption{Accretion rates for different EOSs and neutron star densities, all for black holes with puncture mass ${\mathcal M} = 10^{-6} M_\odot$.  The rest-mass densities $\rho_{0*}$ refer to those observed by a local asymptotic observer, and are very similar to the central density of the corresponding neutron star in the absence of a black hole. 
    For each EOS and central density we list, all in units of solar masses, the resulting black-hole gravitational mass $M_{\rm BH}$, the total gravitational mass $M$, the accretion eigenvalue $\lambda$ (\ref{lambda}), the value of $\dot M_0^* / M_{\rm BH}^2$ from (\ref{m_dot}), the resulting analytical accretion rate $\dot M_0^*$ as measured by a local asymptotic observer, this observer's value of the lapse $\alpha_*$ and the corresponding analytical accretion rate as measured by an observer at infinity $\alpha_* \dot M_0^*$ (see \ref{translate_m_dot}), as well as the numerical accretion rate $\dot M_0$ as computed from (\ref{m_dot_numerical}).   For the MS1 EOS and for the smaller central densities the analytical approach of Section \ref{sec:bondi} did not yield solutions; see text for details.
    }
    \label{tab:results}
\end{table*}

\begin{figure}
    \centering
    \includegraphics[width = 0.45 \textwidth]{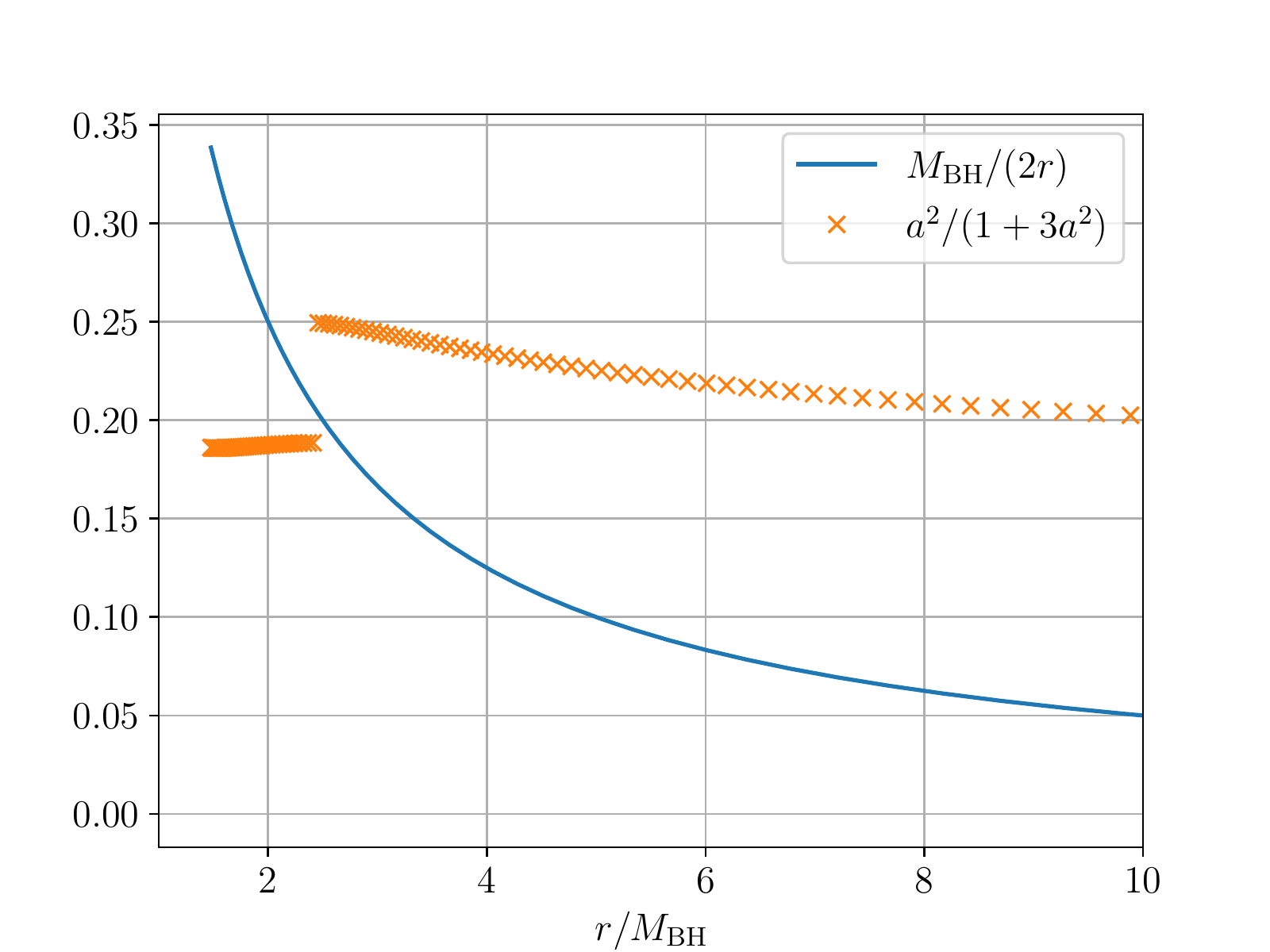}
    \caption{Numerical profiles of the last two terms in the condition (\ref{crit_condition}) for a critical point, namely $M_{\rm BH}/(2r)$ and $a^2 / (1 + 3 a^2)$, for the MS1 EOS with $\rho_0 = 7.5 \times 10^{-4} M_\odot^{-2}$ at coordinate times $t = 1.4 \times 10^3 M_{\rm BH}$, after the evolution has relaxed into stationary equilibrium.  The crosses show grid points used in our simulation.  The discontinuity in the sound speed $a$ prevents an equality of the two terms in this case, so that the analytical procedure of Section \ref{sec:bondi} does not yield a critical point (compare the discussion at the end of Section \ref{sec:bondi}).}
    \label{fig:crit_point}
\end{figure}

\begin{figure*}
    \centering
    \includegraphics[width = 0.45 \textwidth]{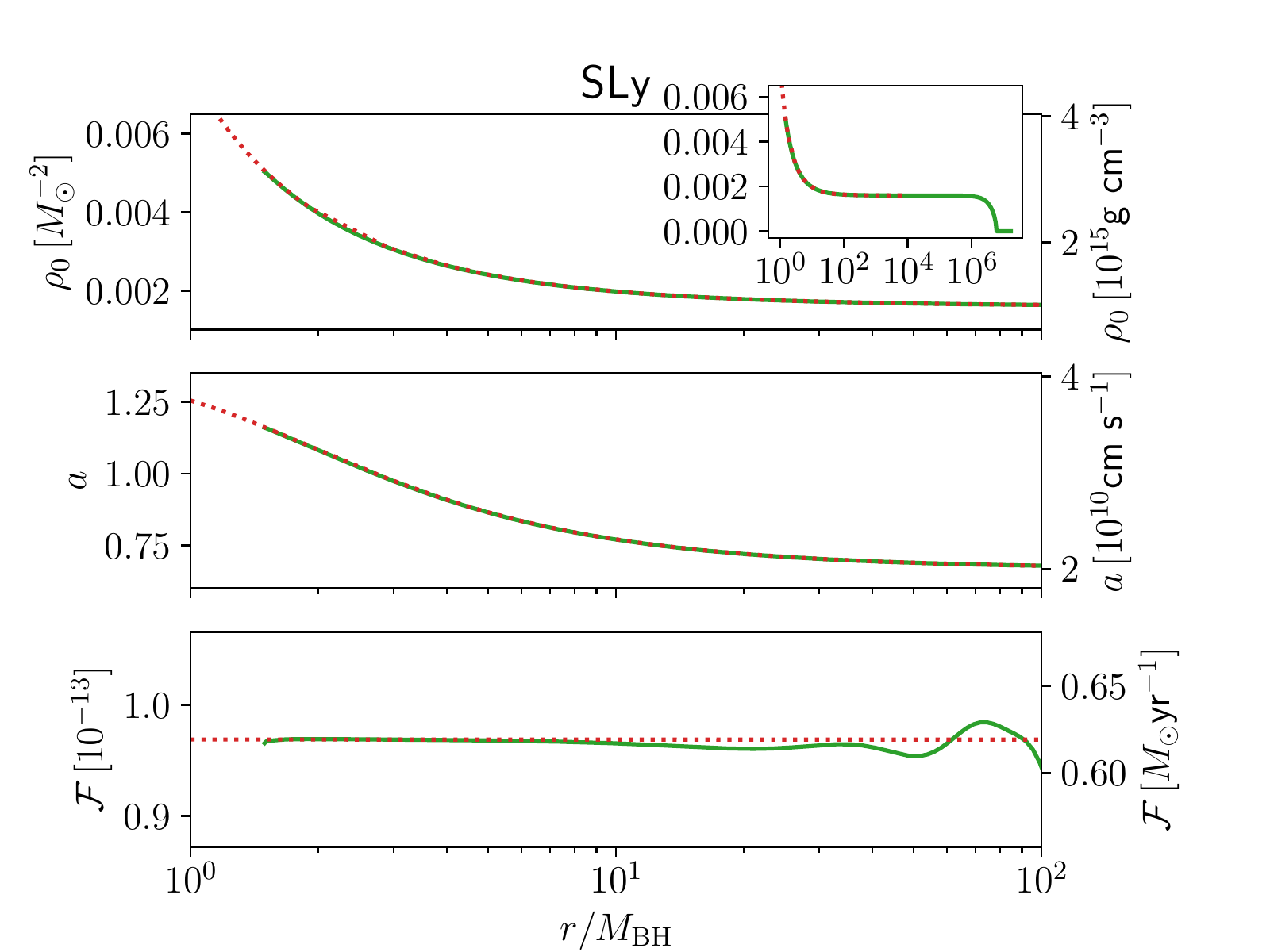}
    \includegraphics[width = 0.45 \textwidth]{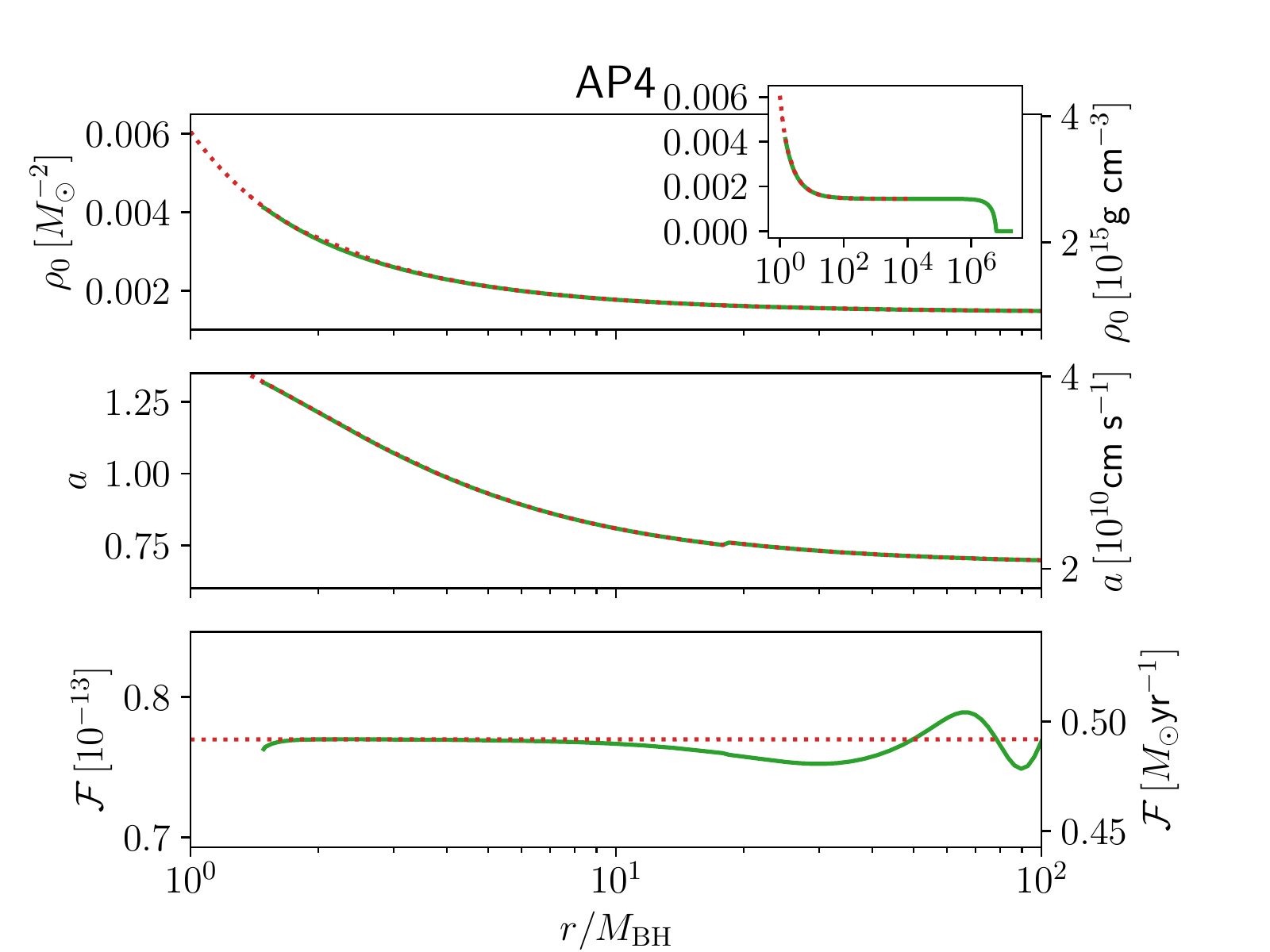}     
    \caption{Analytical and numerical profiles of accretion flow onto a black hole at the center of a neutron star.  On the left we show results for the SLy EOS, starting with a central rest-mass density of $\rho_{0*} = 0.0016 \, M_\odot^{-2} = 9.92 \times 10^{14} \, \mbox{g cm}^{-3}$, while on the right we show results for the AP4 EOS with $\rho_{0*} = 0.00145 \, M_\odot^{-2} = 8.99 \times 10^{14} \, \mbox{g cm}^{-3}$.  The top panels show rest-mass densities, the middle panels the sound speeds, and the bottom panels the flux ${\mathcal F}$, all as functions of areal radius $r$.   In each panel the dotted line represents analytical results from the relativistic Bondi formalism, while the solid lines represent numerical snapshots at coordinate times $t = 1.4 \times 10^3 M_{\rm BH}$, after the solution has settled down to an equilibrium solution in a region around the black hole.  The small insets show the rest-mass densities in the entire star. The oscillations in the flux at large distances from the black hole emerge in the wake of the outgoing shock wave that is triggered by the transition from our initial data to the steady-state accretion solution. }
    \label{fig:example}
\end{figure*}

We perform numerical simulations for the EOSs listed in Table \ref{table:PWP}.  For our analytical treatment, we adopt as the central density of the unperturbed neutron star a value just below that of the maximum mass configuration, as well as some smaller densities, for each EOS (see Table \ref{tab:results}).  We then compute the analytical accretion rates from (\ref{m_dot}).  We compare these rates to our numerical simulation values from (\ref{m_dot_numerical}), adopting neutron star models with the central densities in the absence of the black hole identical to those chosen above.  We provide details of all our results in Table \ref{tab:results}; in particular we list the analytical and numerical values for the accretion rates as measured by an observer at infinity.  

As we discussed in Section \ref{sec:bondi}, the analytical approach described there does not yield analytical solutions for the MS1 EOS for our smaller central densities.   Recall that this approach relies on identifying a critical point defined by equality between the three terms in eq.~(\ref{crit_condition}).  The first of these three points, $u_s$, is a gauge-dependent quantity, but the last two terms are gauge-invariant.  In Fig.~\ref{fig:crit_point} we therefore show  numerical profiles of these two terms for the MS1 EOS and $\rho_0 = 7.5 \times 10^{-4} M_\odot^{-2} = 4.65 \times 10^{14} \mbox{g cm}^{-3}$ at a sufficiently late time for the evolution to have settled into a stationary equilibrium solution close to the black hole.   We see that the difference between the two terms, $M_{\rm BH} / (2 r) - a^2 / (1 - 3 a^2)$, does indeed change sign; however, this difference does not have a root because of the discontinuity of $a$, as we had discussed in Section \ref{sec:bondi}.   This discontinuity, and hence the absence of a point at which equality in (\ref{crit_condition}) holds, is an artifact of the PWP representation of the EOS.   We note, however,
that the numerical solution with its finite grid effectively interpolates between the discontinuity, thereby passing through a critical point and achieving a smooth flow and well-defined accretion rate, as indicated in Table \ref{tab:results}.

In the last column of Table \ref{tab:results} we list the ratios $\dot M_0 / M_{\rm BH}^2$ and note that, for each EOS, these values depend only weakly on the central density or, equivalently, the mass of the neutron star host. Even between different EOSs these values do not vary significantly. We may therefore approximate the accretion rate, for any of the EOSs and central densities considered here, as
\begin{equation} \label{m_dot_summary_1}
    \dot M_0 \simeq \chi \, ( M_{\rm BH} / M_{\odot} )^2
\end{equation}
where, within about 30\% or so, $\chi \simeq 0.06$.  Using $M_\odot = 4.9 \times 10^{-6}\, \mbox{s} = 1.6 \times 10^{-13}\, \mbox{yr}$ we may write (\ref{m_dot_summary_1}) as
\begin{equation} \label{M_dot_summary_2}
    \dot M_0 \simeq 4.0 \times 10^{-9} \, \frac{M_{\odot}}{\mbox{yr}} \, \left( \frac{M_{\rm BH}}{10^{-10} M_\odot} \right)^2,
\end{equation}
which is just slightly larger than the minimum accretion rate reported in \cite{BauS21} (where it was computed under the assumption of single Gamma-law EOSs).\footnote{Note that \cite{BauS21} provided estimates for the rate of gravitational mass-energy accretion, which, during the quasi-stationary Bondi accretion phase, is slightly larger than that for the accretion of rest mass (see Table III in \cite{RicBS21b}).}

We also show examples of accretion profiles, computed both numerically and analytically, in Fig.~\ref{fig:example}.  We note that for many of the examples that we considered these profiles feature superluminal sound speeds $a > c$ in regions close to the black hole.  This behavior is not unexpected; as discussed in RBS, it is unavoidable when $\Gamma \geq 3$, and will also occur for softer EOS with $2 < \Gamma < 3$ if the asymptotic densities are sufficiently high.  Ref.~\cite{EasL19} also noted the appearance of superluminal sound speeds for some EOSs, and artificially adjusted those EOSs in the corresponding high-density regimes to ensure that $a < c$.  We instead allow the sound speed to exceed the speed of light, both in our analytical and numerical treatments.\footnote{In the latter, the sound speed is needed in the approximate Riemann solver.  Transforming the sound speed from the fluid frame to the coordinate frame involves taking the square root of a number that may become negative if $a > c$.  In order to prevent the code from crashing we tried different approaches, including setting this number artificially to zero.  Comparing these different approaches revealed very little difference in our results.}  The possibility of the sound speed becoming superluminal in ultradense matter has certainly been discussed in the past (see, e.g., \cite{BluR68}).  While the equations admit such solutions without breaking down, there are strong arguments for rejecting such behavior on causality grounds, as it violates a basic principle of relativity, as emphasized by \cite{EllMM07}.

The appearance of superluminal sound speeds may be a consequence of either the underlying EOSs or their PWP representation, of course.  To justify some EOSs and their PWP fits for treatments involving stable neutron stars, it is sometimes argued that the sound speed at the stellar core remains subluminal even for the maximum-mass configuration. However, the solutions that we present here provide examples of stationary equilibrium solutions in which the densities significantly exceed superluminal values for such configurations containing small black holes, even outside black-hole horizons. These solutions provide motivation for constructing nuclear EOSs and PWP fits that do not exhibit unphysical superluminal behavior even at these high densities.

\section{Summary}
\label{sec:summary}

We generalize relativistic Bondi solutions describing accretion onto Schwarzschild black holes to allow for realistic, nuclear EOSs approximated by PWPs. In most cases, these solutions can be constructed by identifying a critical point in the accretion flow, as for single Gamma-law EOSs. In just a few cases, however, we found that the discontinuities in the sound speeds, which result from the PWP approximation of the EOS, prevent the identification of the critical point in this approach. However, our time-dependent numerical simulations encounter no problems even for these cases and relax to stationary flows for all cases considered.  We apply our analytical solutions to model accretion onto black holes harbored inside neutron stars, and find excellent agreement with the numerical simulations of this scenario. The accretion rates are all very close to a nearly universal minimum accretion rate (see RBS); they depend primarily on the black hole mass, and only weakly on the EOS and the neutron star properties.  Ignoring the small differences in the accretion rates for rest mass and gravitational mass (see \cite{RicBS21b,BauS21}) we may also integrate (\ref{M_dot_summary_2}) to obtain the neutron star's survival time
\begin{equation} \label{t_max}
    t_{\rm surv} \simeq \frac{M_\odot}{\chi} \left( \frac{M_\odot}{M_0} \right)
    \simeq 8.2 \times 10^{5} \, \mbox{s} \, \left( \frac{10^{-10} M_\odot}{M_0} \right),
\end{equation}
where $M_0$ is the initial black hole mass, and where we have adopted $\chi \simeq 0.06$ in the last equality.  The survival time (\ref{t_max}) is slightly smaller, but close to the maximum survival time reported by \cite{BauS21}, $t_{\rm max} \simeq 1.0 \times 10^{6} \, \mbox{s} \, (10^{-10} M_\odot / M_0)$.  This timescale has been invoked to set constraints on the masses of primordial black holes \cite{CapPT13,GenST20} (but see also \cite{MonFVSH19,KaiNSY21}).

Finally, we caution that many of these solutions feature superluminal sound speeds, and emphasize the requirement that any viable EOS and its PWP representation must avoid this unphysical behavior at the supranuclear densities encountered here.

\acknowledgments

It is a pleasure to thank Charles Gammie, Chloe Richards, Lunan Sun, and Antonios Tsokaros for numerous helpful conversations.  SCS acknowledges support through an undergraduate research fellowship at Bowdoin College. This work was supported in part by National Science Foundation (NSF) grants PHY-1707526 and PHY-2010394 to Bowdoin College, and NSF grants PHY-1662211 and PHY-2006066 and National Aeronautics and Space Administration (NASA) grant  80NSSC17K0070 to the University of Illinois at Urbana-Champaign.

%


\end{document}